\documentstyle[epsfig,a4p,12pt,rotating]{article}
\begin{document}
%
%
\newcommand{\PPEnum}    {CERN-EP/99-084}
\newcommand{\PRnum}     {OPAL PR-281}
\newcommand{\PNnum}     {OPAL Physics Note PN-xxx}
\newcommand{\TNnum}     {OPAL Technical Note TN-xxx}
\newcommand{\Date}      {21 June 1999}
\newcommand{\Author}    {M.~J.~Oreglia}
\newcommand{\MailAddr}  {Mark.Oreglia@cern.ch}
\newcommand{\EdBoard}   {S. Arcelli, T. Junk, K. Nagai,A. Quadt}
\newcommand{\DraftVer}  {DRAFT 4.7}
\newcommand{\DraftDate} {\Date}
\newcommand{\TimeLimit} {Monday, 16 June 1999, 17h00 Geneva time}

\def\toprule{\noalign{\hrule \medskip}}
\def\midrule{\noalign{\medskip\hrule }}
\def\botrule{\noalign{\medskip\hrule }}
\setlength{\parskip}{\medskipamount}

\newcommand{\TPONE}{$\theta_{\gamma 1}$}
\newcommand{\TPTWO}{$\theta_{\gamma 2}$}
\newcommand{\CTPM}{$\cos\theta_{\mrm miss}$}
\newcommand{\ECV} {$E_{\mrm{excess}}$}
\newcommand{\NPHO} {$\mrm N_{\gamma} > 1$}

\newcommand{\mgg} {$M_{\gamma \gamma}$}
\newcommand{\mdip} {M_{\gamma \gamma}}
\newcommand{\Stogg} {${\mathrm S} \ra \gamma \gamma$}
\newcommand{\epem}{{\mathrm e}^+ {\mathrm e}^-}
\newcommand{\tptm}{{\tau}^+ {\tau}^-}
\newcommand{\mpmm}{{\mu}^+ {\mu}^-}

\newcommand{\qqbar}{{\mathrm q}\bar{\mathrm q}}
\newcommand{\nn}{\nu \bar{\nu}}
\newcommand{\nunu}{\nu \bar{\nu}}
\newcommand{\mumu}{\mu^+ \mu^-}
\newcommand{\ellell}{\ell^+ \ell^-}
\newcommand{\MZ}{M_{\mathrm Z}}
\newcommand{\MH}{M_{\mathrm H}}
\newcommand{\MX} {M_{\mathrm{X}}}
\newcommand{\MY} {M_{\mathrm{Y}}}

\newcommand {\Hboson}        {{\mathrm h}^{0}}
\newcommand {\Hzero}         {${\mathrm h}^{0}$}
\newcommand {\Zboson}        {{\mathrm Z}^{0}}
\newcommand {\Zzero}         {${\mathrm Z}^{0}$}
\newcommand {\Azero}         {${\mathrm A}^{0}$}
\newcommand {\Wpm}           {{\mathrm W}^{\pm}}
\newcommand {\allqq}         {\sum_{q \neq t} q \bar{q}}
\newcommand {\mixang}        {\theta _{\mathrm {mix}}}
\newcommand {\thacop}        {\theta _{\mathrm {Acop}}}
\newcommand {\cosjet}        {\cos\thejet}
\newcommand {\costhr}        {\cos\thethr}
\newcommand{\epair}    {\mbox{${\mathrm e}^+{\mathrm e}^-$}}
\newcommand{\mupair}   {\mbox{$\mu^+\mu^-$}}
\newcommand{\taupair}  {\mbox{$\tau^+\tau^-$}}
\newcommand{\qpair}    {\mbox{${\mathrm q}\overline{\mathrm q}$}}
\newcommand{\ff}       {{\mathrm f} \bar{\mathrm f}}
\newcommand{\gaga}     {\gamma\gamma}
\newcommand{\WW}       {{\mathrm W}^+{\mathrm W}^-}
\newcommand{\eeee}     {\mbox{\epair\epair}}
\newcommand{\eemumu}   {\mbox{\epair\mupair}}
\newcommand{\eetautau} {\mbox{\epair\taupair}}
\newcommand{\eeqq}     {\mbox{\epair\qpair}}
\newcommand{\eeff}     {\mbox{$\mathrm e^+e^- f \bar{\mathrm f}$}}
\newcommand{\llnunu}   {\mbox{\lpair\nu\nubar}}
\newcommand{\lnuqq}    {\mbox{\lept\nubar\qpair}}
\newcommand{\zee}      {\mbox{Zee}}
\newcommand{\wenu}     {\mbox{We$\nu$}}

\newcommand{\el}       {\mbox{${\mathrm e}^-$}}
\newcommand{\selem}    {\mbox{$\tilde{\mathrm e}^-$}}
\newcommand{\smum}     {\mbox{$\tilde\mu^-$}}
\newcommand{\staum}    {\mbox{$\tilde\tau^-$}}
\newcommand{\slept}    {\mbox{$\tilde{\ell}^\pm$}}
\newcommand{\sleptm}   {\mbox{$\tilde{\ell}^-$}}
%
%
\newcommand{\zz}        {\mbox{$|z_0|$}}
\newcommand{\dz}        {\mbox{$|d_0|$}}
\newcommand{\sint}      {\mbox{$\sin\theta$}}
\newcommand{\cost}      {\mbox{$\cos\theta$}}
\newcommand{\mcost}     {\mbox{$|\cos\theta|$}}
\newcommand{\dedx}      {\mbox{$dE/dx$}}
\newcommand{\wdedx}     {\mbox{$W_{dE/dx}$}}
\newcommand{\xe}        {\mbox{$x_{\rm E}$}}
\newcommand{\xgam}      {x_{\gamma}}
\newcommand{\Mrec}      {M_{\mrm{recoil}}}
\newcommand{\Lgg}       {\mbox{${\cal L}(\gaga)$}}
\newcommand{\ggjj}{\mbox{$\gaga\;{\rm jet-jet}$ }}
\newcommand{\ggqq}{\mbox{$\gaga {\rm q\bar{q}}$ }}
\newcommand{\ggll}{\mbox{$\gaga\ell^{+}\ell^{-}$}}
\newcommand{\ggnn}{\mbox{$\gaga\nu\bar{\nu}$}}
\newcommand{\eeeell}{\mbox{\epair$\rightarrow$\epair\lpair}}
\newcommand{\eell}{\mbox{\epair\lpair}}
\newcommand{\llgam}{\mbox{$\ell\ell(\gamma)$}}
\newcommand{\nunugam}{\mbox{$\nu\bar{\nu}\gaga$}}
\newcommand{\acope}{\mbox{$\Delta\phi_{\mathrm{EE}}$}}
\newcommand{\nee}{\mbox{N$_{\mathrm{EE}}$}}
\newcommand{\eesum}{\mbox{$\Sigma_{\mathrm{EE}}$}}
\newcommand{\acoph}{\mbox{$\Delta\phi_{\mathrm{HCAL}}$}}
\newcommand {\mm}         {\mu^+ \mu^-}
\newcommand {\emu}        {\mathrm{e}^{\pm} \mu^{\mp}}
\newcommand {\et}         {\mathrm{e}^{\pm} \tau^{\mp}}
\newcommand {\mt}         {\mu^{\pm} \tau^{\mp}}
\newcommand {\lemu}       {\ell=\mathrm{e},\mu}
\newcommand {\Zz}         {\mbox{${\mathrm{Z}^0}$}}
%

\newcommand{\gsim}{\;\raisebox{-0.9ex}
           {$\textstyle\stackrel{\textstyle >}{\sim}$}\;}

\newcommand{\degree}    {^\circ}
%
\newcommand{\roots}     {\sqrt{s}}
\newcommand{\Ecm}         {\mbox{$E_{\mathrm{cm}}$}}
\newcommand{\Ebeam}       {E_{\mathrm{beam}}} 
\newcommand{\ipb}         {\mbox{pb$^{-1}$}}
%
%
\newcommand{\thmiss}    { \theta_{miss} }
\newcommand{\cosmiss}   {| \cos \thmiss |}
%
%
\newcommand{\Evis}      {\mbox{$E_{\mathrm{vis}}$}}
\newcommand{\Rvis}      {\mbox{$R_{\mathrm{vis}}$}}
\newcommand{\Mvis}      {\mbox{$M_{\mathrm{vis}}$}}
\newcommand{\Rbal}      {\mbox{$R_{\mathrm{bal}}$}}
%
%
%
\newcommand{\PhysLett}  {Phys.~Lett.}
\newcommand{\PRL}       {Phys.~Rev.\ Lett.}
\newcommand{\PhysRep}   {Phys.~Rep.}
\newcommand{\PhysRev}   {Phys.~Rev.}
\newcommand{\NPhys}     {Nucl.~Phys.}
\def\NIM                {\mbox{Nucl. Instr. Meth.}}
\newcommand{\NIMA}[1]   {\NIM\ {\bf A{#1}}}
\newcommand{\IEEENS}    {IEEE Trans.\ Nucl.~Sci.}
\newcommand{\ZPhysC}[1]    {Z. Phys. {\bf C#1}}
\newcommand{\EurPhysC}[1]    {Eur. Phys. J. {\bf C#1}}
\newcommand{\PhysLettB}[1] {Phys. Lett. {\bf B#1}}
\newcommand{\CPC}[1]      {Comp.\ Phys.\ Comm.\ {\bf #1}}
\def\etal{\mbox{{\it et al.}}}
%
%
\newcommand{\OPALColl}  {OPAL Collab.}
\newcommand{\JADEColl}  {JADE Collab.}
%
\newcommand{\onecol}[2] {\multicolumn{1}{#1}{#2}}
\newcommand{\colcen}[1] {\multicolumn{1}{|c|}{#1}}
\newcommand{\ra}        {\rightarrow}   
\newcommand{\ov}        {\overline}   
\def\mrm       {\mathrm}
\newcommand {\downto}
        {\mbox{ \begin{picture}(14,10)
                   \put(0,10){\line(0,-1){5.0}}
                   \put(2,5){\oval(4,4)[bl]}
                   \put(1,0){\makebox(0,0)[bl]{$\rightarrow$}}
                \end{picture} }}


\begin{titlepage}
\begin{center}{\large   EUROPEAN LABORATORY FOR PARTICLE PHYSICS
}\end{center}\bigskip
\begin{flushright}
       \PPEnum  \\ \DraftDate
\end{flushright}
\bigskip\bigskip\bigskip\bigskip\bigskip
%
%
\begin{center}{\huge\bf\boldmath Search for Higgs Bosons \\
                        and Other Massive States \\
                        Decaying into Two Photons \\
\vspace{3pt}
                        in $\epem$ Collisions at 189 GeV }
\end{center}\bigskip\bigskip
\begin{center}{\LARGE The OPAL Collaboration
}\end{center}\bigskip\bigskip
%
%
\bigskip\begin{center}{\large  Abstract}\end{center}
A search is described for the generic process 
$\epem\ra {\mrm X Y}$,
where X is a neutral heavy scalar boson decaying into a pair of photons,
and Y is a neutral heavy boson (scalar or vector) decaying into 
a fermion pair. 
The search is motivated mainly by the cases where
either X, or both X and Y, are Higgs bosons. 
In particular, we investigate
the case where X is the Standard Model Higgs boson and Y the \Zzero\ boson.
Other models with enhanced Higgs boson decay couplings to photon pairs
are also considered. 
The present search combines the data set collected by
the OPAL collaboration at 189 GeV collider energy,
having an integrated luminosity of 182.6~\ipb, 
with data samples collected at lower energies. 
The search results have been used to put
95\% confidence level bounds, as functions of the mass $\MX$, on the product
of the cross-section and the relevant branching ratios, both in a model
independent manner and for the particular models considered.

\bigskip\bigskip\bigskip\bigskip

\begin{center}{\large
(Accepted by Physics Letters B) \\
}\end{center}

\end{titlepage}

\begin{center}{\Large        The OPAL Collaboration
}\end{center}\bigskip
\begin{center}{
G.\thinspace Abbiendi$^{  2}$,
K.\thinspace Ackerstaff$^{  8}$,
G.\thinspace Alexander$^{ 23}$,
J.\thinspace Allison$^{ 16}$,
N.\thinspace Altekamp$^{  5}$,
K.J.\thinspace Anderson$^{  9}$,
S.\thinspace Anderson$^{ 12}$,
S.\thinspace Arcelli$^{ 17}$,
S.\thinspace Asai$^{ 24}$,
S.F.\thinspace Ashby$^{  1}$
% ...
}\end{center}\bigskip
\bigskip
$^{  1}$School of Physics and Astronomy, University of Birmingham,
Birmingham B15 2TT, UK
\newline
$^{  2}$Dipartimento di Fisica dell' Universit\`a di Bologna and INFN,
I-40126 Bologna, Italy
\newline
% ...
$^{  5}$Cavendish Laboratory, Cambridge CB3 0HE, UK
\newline
% ...
$^{  8}$CERN, European Organisation for Particle Physics,
CH-1211 Geneva 23, Switzerland
\newline
$^{  9}$Enrico Fermi Institute and Department of Physics,
University of Chicago, Chicago IL 60637, USA
\newline
% ...
$^{ 12}$Indiana University, Department of Physics,
Swain Hall West 117, Bloomington IN 47405, USA
\newline
% ...
$^{ 16}$Department of Physics, Schuster Laboratory, The University,
Manchester M13 9PL, UK
\newline
$^{ 17}$Department of Physics, University of Maryland,
College Park, MD 20742, USA
\newline
% ...
$^{ 23}$Department of Physics and Astronomy, Tel Aviv University,
Tel Aviv 69978, Israel
\newline
$^{ 24}$International Centre for Elementary Particle Physics and
Department of Physics, University of Tokyo, Tokyo 113-0033, and
Kobe University, Kobe 657-8501, Japan
\newline
% ...
$^{  d}$ on leave of absence from the University of Freiburg
\newline
$^{  e}$ and University of Mining and Metallurgy, Cracow
\newline
$^{  f}$ and Heisenberg Fellow
\newline
% ...
\newline

\newpage
\section{Introduction}
\label{sec:intro}

We present a search for a di-photon
resonance produced in $\epem$ collisions at LEP.
The data were taken by the OPAL detector at
centre-of-mass energies \Ecm\ up to 189~GeV. 
The search is sensitive to the process 
$\epem \ra \mrm X Y$, with $\mrm X \ra \gaga$ and $\mrm Y \ra \ff $,
where $\ff$ may be quarks, charged leptons, or a neutrino pair.
In a Standard Model scenario, Y is a \Zzero\
and X is a Higgs boson decaying into two photons. 
A more general search
is achieved by removing the restriction that Y is a \Zzero.

In the minimal Standard Model, the single Higgs boson
can decay into two photons via a quark- or W-boson loop~\cite{HBR}.
The rate is too small for observation at existing accelerators even for a 
kinematically accessible Higgs boson, but
other theoretical models can accommodate large $\Hboson \ra \gaga$
branching ratios~\cite{Hagiwara}. 
Throughout this paper, ``\Hzero'' refers to a neutral CP-even scalar where 
non-minimal Higgs sector models are discussed. 
Particularly interesting are non-minimal Higgs sectors
wherein some Higgs components
couple only to bosons~\cite{Fermiophobic}.
This class of ``fermiophobic'' Higgs models includes the 
``Bosonic'' Higgs model~\cite{Bosonic}, 
and Type I Two-Higgs Doublet models with fermiophobic couplings~\cite{TypeI}.
In Higgs triplet models~\cite{Gunion}, 
the particles formed from the triplet fields are fermiophobic.

There are existing limits on the production of a di-photon resonance 
which couples 
to the \Zzero. 
Using data taken up to \Ecm=183~GeV, OPAL has set upper 
limits on the branching ratio $\Hboson \ra \gaga$ for masses 
up to 92~GeV~\cite{183paper,OPAL_HGG}
and obtained a 95\% confidence level (CL) lower mass limit of 90.0~GeV  for
a fermiophobic Higgs scalar. 
%
Other collaborations~\cite{D0paper,Delphi1999} have recently reported
limits on photonic Higgs boson decays.
The lower mass region (\mgg $<$ 60~GeV) has been searched previously
using data from LEP-I~\cite{OPAL_ggjj_1,LowMgg,other_gg}.

\section{Data and Monte Carlo Samples}
%
The analysis is performed on the data collected 
with the OPAL detector~\cite{detector}
during the 1998 LEP run.
The data sample corresponds to an integrated luminosity of $182.6\pm0.8$ \ipb\
collected at a luminosity-weighted \Ecm\ of $188.63 \pm 0.04$~GeV. 

To assess the sensitivity of the analysis to signals, two production models are
considered: the Standard Model process $\epem \ra \Hboson\Zboson$, 
and Two Higgs Doublet models (2HDM) for $\epem \ra \Hboson \mrm A^0$. 
The process 
$\epem \ra \Hboson\Zboson$, $\Hboson \ra \gaga$ 
was simulated for each \Zzero\ decay channel
using the HZHA generator~\cite{HZHA}. 
For the general search, mass grids were generated using the 
$\epem \ra \Hboson \Zboson$ and
$\epem \ra \Hboson \mrm A^0$
processes as models 
for $\epem \ra {\mrm XY},~\mrm{X} \ra \gamma\gamma,~\mrm{Y} \ra \ff$.

The dominant background to this search arises from the
emission of two energetic initial state radiation (ISR) photons in hadronic events from
$\epem \ra (\gamma/{\rm Z)^{\ast} \ra {\rm q\bar{q}}} $.
This process was simulated using
the KK2f generator
using CEEX~\cite{CEEX} ISR modelling, and
with the set of hadronization parameters described in reference \cite{jtparams}.
Other Standard Model backgrounds, particularly those from 4-fermion processes,
primarily affect the leptonic and missing energy modes of the search.
Four-fermion
processes were
modelled using the Vermaseren \cite{VERMASEREN} and grc4f \cite{grc4f} 
generators implemented in the KORALW~\cite{KORALW} Monte Carlo program. 
The programs 
BHWIDE~\cite{BHWIDE} and 
TEEGG~\cite{TEEGG} were employed
to model the s- and t-channel backgrounds from Bhabha scattering. 
The processes $\epem \ra \ellell$ with $\ell \equiv \mu , \tau$ 
were simulated using KORALZ~\cite{KORALZ}. 
The KORALZ program was also used to 
generate events of the type $\epem\ra\nu\ov{\nu}\gamma(\gamma)$.
The process $\epem \ra \gaga$ was simulated using the
RADCOR generator~\cite{RADCOR}.
Simulated events were processed using the full
OPAL detector Monte Carlo~\cite{GOPAL} and analyzed in the 
same manner as the data.

\section{Event Selection}

In our searches for the generic process 
$\epem \ra {\mrm XY}$,
we consider three event topologies
which are motivated mainly by the particular case where 
X is a generic Higgs boson decaying into a pair of photons, 
and Y is the \Zzero\ boson decaying either into
(1) a $\qqbar$ pair, or 
(2) a pair of oppositely charged leptons, or
(3) a $\nn$ pair.
The search topologies are therefore:

\begin{itemize}

\item Two energetic photons recoiling against a hadronic system.  

\item Two energetic photons produced in association with charged leptons.

\item Two energetic photons and no other significant detector activity.
\end{itemize}

In the \Hzero\Zzero\ search, the mass recoiling from the di-photon system
is required to be consistent with the \Zzero\ mass for all topologies,
while in the general search this condition is not required.
A background common to all search modes arises from events with two
visible ISR photons, resulting in an on-shell 
\Zzero\ recoiling from a di-photon system.

The selection criteria employed in this search are very similar to
those described in reference~\cite{183paper}.
For all topologies, charged tracks (CT) and unassociated 
electromagnetic calorimeter (EC) clusters are 
required to satisfy the criteria defined in reference \cite{CTSEL}.
``Unassociated'' EC clusters are defined by the requirement that
no charged tracks point to the cluster.
For each channel, preselection cuts are 
applied which employ the following measured quantities:
\begin{itemize}
\item $\Evis$ and $\vec{p}_{\mrm{vis}}$: the scalar and vector sums
      of charged track momenta, unassociated EC and 
      unassociated hadron calorimeter cluster energies.
\item $\Rvis \equiv \frac{\mbox{\Evis}}{\Ecm}$.

\item Visible momentum along the beam direction: 
      $|\Sigma~p_{z}^{\mrm{vis}}|$,
      where all tracks and unassociated clusters are summed over.
\end{itemize}

\subsection{Photon Pair Identification}
\label{s:photid}

After channel-dependent preliminary cuts based on data quality
and rudimentary 
event topology (described in the next sections), 
events are required to have a photon pair satisfying several criteria. 
Photon identification
is accomplished by identifying clusters
in the electromagnetic calorimeter.
These EC clusters are combined with the information from the tracking
detectors to identify photon candidates
if the lateral spread of the 
clusters satisfies the criteria described in 
reference \cite{OPAL_HGG}.
The photon detection efficiency is increased
by including photon conversions into \epair\ pairs
using the methods described in reference~\cite{183paper}.

\begin{figure}[!p]
        \vspace{0.8cm}
        \begin{center}
           \resizebox{\linewidth}{!}{\includegraphics{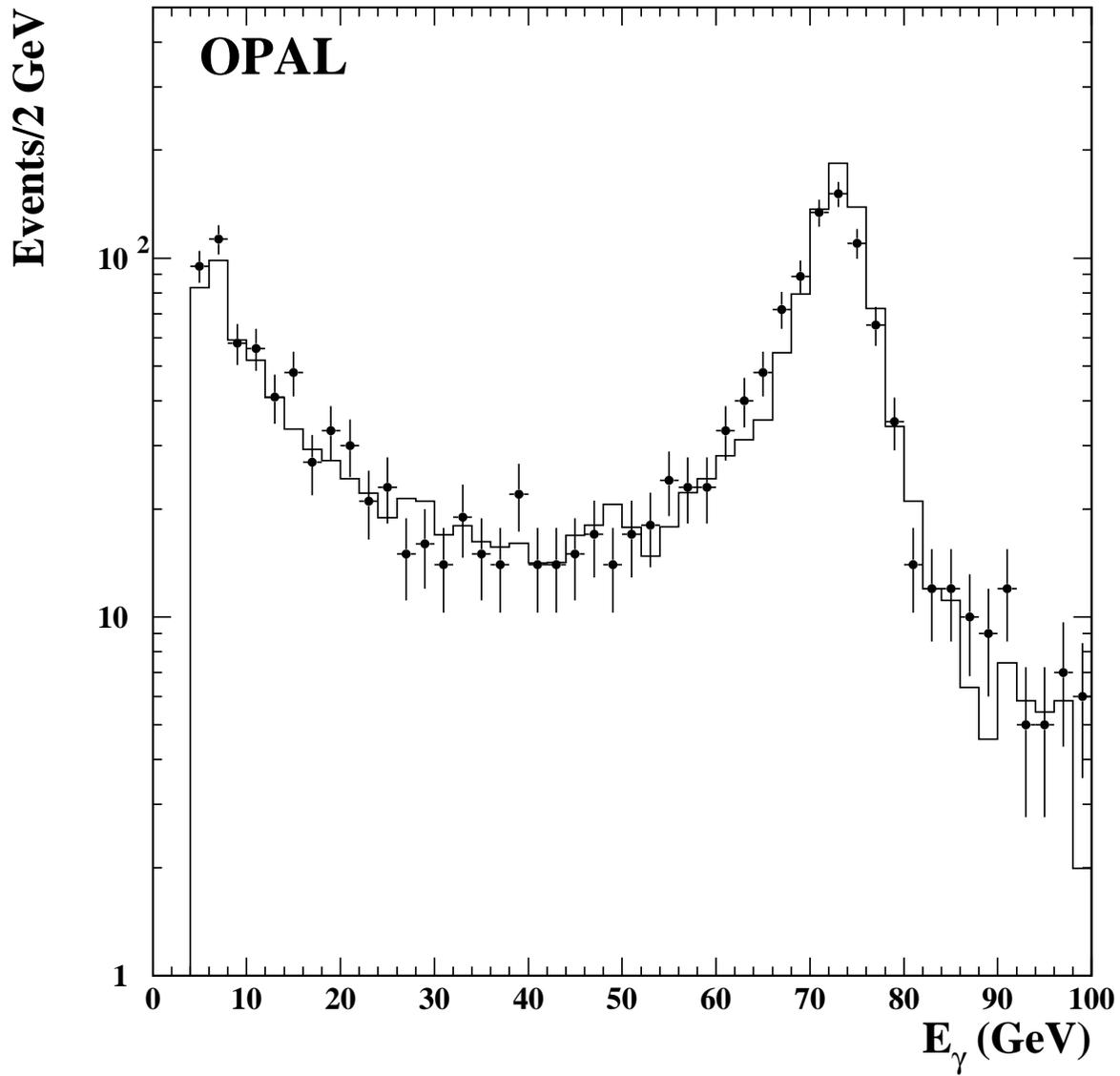} }
        \caption[CFISR]{    
                  Energy distribution of highest-energy photon in the
                  hadronic search channel.
                  Data are shown as points with error bars.
                  Background simulation is shown as a histogram.
        \label{CFISR} }
        \end{center}
    \end{figure}

The most significant background to all search channels 
are processes having two ISR photons. 
Photons from ISR are peaked along the beam direction,
hence cutting on the photon polar angle
$|\cos(\theta_{\gamma})| < 0.875 $ 
is very effective in 
reducing the background acceptances without significantly decreasing
the efficiencies for potential signals.
Figure~\ref{CFISR} shows the distribution of 
$E_{\gamma1}$, the highest-energy photon, in events having
hadronic activity (criterion A1 described below), 
where at least one photon has
$E_{\gamma} > 0.05 \times E_{\mrm beam}$ and 
$|\cos(\theta_{\gamma})| < 0.875 $. 
Also shown is the simulated Standard Model background.  
The overall number of background photons, their energies, 
and their polar angle distribution (not shown) describe the data 
to better than 10\%. 
The photon pair acceptance criteria is thus summarized by the following requirements
on the two highest-energy photons in the event:
\begin{itemize}

 \item The two photon candidates are required to be in the fiducial region
       $|\cos(\theta_{\gamma})| < 0.875$, where
       $\theta_{\gamma}$ is the angle of the photon 
       with respect to the $\mrm e^-$ beam direction.

 \item The higher energy photon is required to have 
       $E_{\gamma}/E_{\mrm beam} > 0.10$ and
       the second-highest-energy photon is required to have 
       $E_{\gamma}/E_{\mrm beam} > 0.05$.

\end{itemize}

After the final channel cuts described in the next sections,
there are no events in which more than one photon pair is found.

\subsection{Hadronic Channel}
\label{s:qqgg}
The hadronic channel is characterized by two photons recoiling against a hadronic system.
In addition to double ISR, backgrounds also arise
from radiative $\Zboson\gamma$ events where a decay product of the
\Zzero, such as an isolated $\pi^0$ or $\eta$ meson, mimics a photon, 
or there is an energetic final state radiation (FSR) photon.
In these cases, the recoil mass against the di-photon system will tend to be 
lower than the \Zzero\ mass; therefore, this background can be suppressed by 
requiring a recoil mass consistent with that of the \Zzero.
In the general search for XY $\ra$ $\gaga~+~\mrm{hadrons}$, 
there is no recoil
mass constraint to help suppress backgrounds from fake photons.

The hadronic channel candidate selection is summarized in Table
\ref{T:qq1}. Candidate events are 
required to satisfy the following criteria:
\begin{itemize}
  \item[(A1)] The standard hadronic event preselection
              described in reference~\cite{hadsel}
              with the additional requirements:
    \begin{itemize}
       \item $\Rvis > 0.5$ and $|\Sigma~p_{z}^{\mrm{vis}}| < 0.6 \Ebeam$;
       \item at least 2 electromagnetic clusters 
             with $E/\Ebeam > 0.05$.
    \end{itemize}

   \item[(A2)] The photon pair criteria described in Section~\ref{s:photid}.

   \item[(A3)] To suppress the background from FSR and fake photons,
               the charged tracks and unassociated clusters were 
               grouped into two jets using the Durham scheme \cite{Durham}, 
               excluding the photon candidates. 
               Both photon 
               candidates are then required to satisfy  
               $p_{\mrm T,~jet-\gamma} > 5$~GeV, where 
               $p_{\mrm T,~jet-\gamma}$ was defined as the photon 
               momentum transverse to the axis defined by the closest jet. 

\end{itemize}

In the case of double ISR emission in the hadron channel, the photons tend
to have a large energy difference.
We therefore employ the quantity 
$\Delta E = (E_{\gamma1}-E_{\gamma2})/{\mrm E_{o}}$, where
$E_{\gamma1}$ and $E_{\gamma2}$ are the first- and second-highest-energy photons
in the event,
and ${\mrm E_{o}} = (\mrm{E_{cm}}^{2}-\MZ^{2})/(2\mrm{E_{cm}})$
is the energy of a single photon recoiling from the \Zzero.
%

\begin{itemize}

   \item[(A4)] $\Delta E < 0.5$.

   \item[(A5)]For the $\Hboson\Zboson$ topology, 
                the invariant mass
                recoiling from the di-photon must satisfy 
                $|M_{\mrm recoil} - M_{\mrm Z}| < 20$~GeV. 

\end{itemize}

For the general search topology, where no explicit
recoil mass cut is made, 
16 events are observed, while $17.4\pm1.7$ are expected
from Standard Model backgrounds.
The uncertainty shown is for Monte Carlo statistics 
only~\footnote{Unless otherwise specified, all errors quoted are statistical only.}.
After applying the cut (A5) on the recoil mass, 10 events remain, compared to
an expectation of $9.0\pm1.3$ events.
The efficiencies for this analysis to accept events 
for Higgs masses of 30 to 100~GeV are shown in Table~\ref{t:hzeff}.

\subsection{Charged Lepton Channel}
\label{s:llgg}

This channel searches for events in the $\gaga \ellell $ final state.
Even in the case of $\ell = \tau$, this channel has a very clean signature,
and therefore only one selection procedure is required for 
the $\mrm e,\mu$ and $\tau$ channels. 
Charged leptons are identified as low multiplicity jets 
formed from charged tracks and isolated EC clusters.
A high efficiency is maintained for $\tau$ leptons
by allowing single charged tracks to define a ``jet'' without
requiring explicit lepton identification.
This channel is sufficiently free of background to allow
acceptance of events where one of the charged leptons was not
reconstructed or was lost in uninstrumented regions of the detector.
The most serious background comes from 
Bhabha scattering with initial and/or final state radiation. 

The leptonic channel event selection is summarized in Table~\ref{T:ggll}.
Leptonic channel candidates are required to satisfy the following
criteria:

\begin{itemize}
   \item[(B1)] The low multiplicity preselection of reference~\cite{lowmsel} and:
 \begin{itemize}
   \item $\Rvis>0.2$ and $|\Sigma~p_{z}^{\mrm{vis}}|<0.8 \Ebeam$;
   \item number of EC clusters not associated with tracks:  
         $N_{\rm EC} \leq 10$;
   \item number of charged tracks: 
         $1 \leq N_{\rm CT} \leq 7$;
   \item at least 2 electromagnetic clusters 
             with $E/\Ebeam > 0.05$.
  \end{itemize}

  \item[(B2)] The photon pair criteria described in Section~\ref{s:photid}.

  \item[(B3)] For events having only one charged track, require:
     \begin{itemize} 
       \item the track not to be associated with a converted photon;
       \item the track to have momentum satisfying $p>0.2E_{\mrm beam}$;
       \item direction of event missing momentum: 
             $|\cos\theta_{\mrm miss}| > 0.90$. 
     \end{itemize}

  \item[(B4)] For events having two or more charged tracks, the event
              is forced to have 2 jets within the Durham scheme, excluding
              the identified di-photon candidate. 

  \item[(B5)] For the \Hzero\Zzero\ search, the recoil mass to the
                di-photon is required to be consistent with the \Zzero:
              $|\Mrec - \MZ| < 20$~GeV.
\end{itemize} 

For the general search, 20 events survive cuts B1-B4,
compared to $25.6\pm1.6$ expected from Standard Model backgrounds.
After the recoil mass requirement, the number of
observed events is 7, with the background expectation of $8.9\pm1.0$.
The efficiencies for Higgs masses of 30 to 100~GeV are given in 
Table~\ref{t:hzeff}.

\subsection{Missing Energy Channel}
\label{s:nngg}

The missing energy channel is characterized by two photons 
and no other significant detector activity.
An irreducible Standard Model background
is the process $\epem \ra \nu\bar{\nu} \gaga$. 
Other potential backgrounds include 
$\epem \ra \gaga(\gamma)$ and radiative Bhabha scattering with
one or more unobserved electrons. 
These backgrounds tend to produce photons near the beam directions;
therefore, they can be effectively dealt with by the restriction
on the polar angles of the two photons and by requiring
consistency with a di-photon recoiling from a massive object.

The event selection for the missing energy channel is summarized
in Table~\ref{T:ggnn}.
Candidates in the missing energy channel are required to 
satisfy the following criteria:

\begin{itemize}
\item[(C1)] The low multiplicity preselection of reference~\cite{lowmsel} with
            the further requirement that the event satisfy the    
            cosmic ray and beam-wall/beam-gas vetoes
            described in reference \cite{photsel}, and:
    \begin{itemize}
     \item number of EC clusters not associated with tracks:  
         $N_{\rm EC} \leq 4$;
     \item number of charged tracks: 
         $N_{\rm NCT} \leq 3$;
     \item $|\Sigma~p_{z}^{\mrm{vis}}|<0.8 \Ebeam$;
     \item at least 2 electromagnetic clusters 
               with $E/\Ebeam > 0.05$.
    \end{itemize}

    \item[(C2)] The photon pair criteria described in Section~\ref{s:photid}.

    \item[(C3)] Consistency with the hypothesis that the di-photon system
            is recoiling from a massive body:
      \begin{itemize}            
         \item The momentum component of the di-photon system in the plane 
               transverse to the beam axis: $p_T (\gaga)>0.05 E_{\mrm beam}$.
         \item The angle between the two photons in the plane 
               transverse to the beam axis: 
               $|\phi_{\gaga}-180\degree| > 2.5\degree$.
         \item The polar angle of the momentum of the di-photon system: 
               $|\cos\theta_{\gaga}| < 0.966$.
      \end{itemize}

    \item[(C4)] Events are required to have no charged track
            candidates (other than those associated with an identified photon 
            conversion).

    \item[(C5)] Veto on unassociated calorimeter energy: the energy observed 
              in the  electromagnetic
              calorimeter not associated with the 2 photons is required to
              be less than 3~GeV.

    \item[(C6)] For the \Hzero\Zzero\ search, the recoil mass against
                the di-photon is required to be consistent with the \Zzero:
              $|\Mrec - \MZ| < 20$~GeV.
\end{itemize}

The number of events passing the general cuts C1-C5 is 8,
compared to the Standard Model background expectation of $11.2\pm0.5$.
After application of the recoil mass cut (C6), 5 candidates
remain compared to an expectation of $7.1\pm0.3$ events from Standard Model sources. 
%
The efficiencies for Higgs masses from 30 to 100~GeV
are summarized in Table \ref{t:hzeff}.

\subsection{Systematic Errors}

The dominant systematic uncertainty for acceptances 
arises from the photon detection efficiency,
primarily due to the simulation of the photon isolation criteria \cite{OPAL_ggjj_1}.
This uncertainty is estimated to be 3\% of the acceptance 
from comparison of data with Standard Model backgrounds. 
Photon energies and angles are well measured and consequently
lead to a systematic uncertainty on the efficiencies of 0.6\%,
as determined from the measured di-photon recoil mass distribution. 
The systematic error on the integrated luminosity of the data is 0.4\% 
and contributes negligibly to the limits. 
The uncertainty from simulation Monte Carlo statistics
is typically better than 4\%.
From the differences observed in the comparison of data and simulations
of Standard Model backgrounds (particularly the KK2f modelling of ISR),
the systematic uncertainty for backgrounds is taken to be 10\%;
this value is subtracted from the predicted background in the setting of limits.
A systematic error on the photon energy scale is estimated to be 
0.25~GeV for 72~GeV photons 
using the fitted single-photon ISR peak in Figure~\ref{CFISR} 
compared to the expected value based
on the precisely known beam energy and \Zzero\ mass.
This leads to a systematic uncertainty on the di-photon mass
of 0.35~GeV at a mass of 100~GeV.

The background events in the missing energy channel 
include a component from Compton scattering
in the beams, which is modelled by the TEEGG Monte Carlo.  
The photons from this process have a high probability to be found
in the near the cut on polar angle. The photon energy uncertainty 
is rather large (5-9 GeV)
in these regions because of the corrections for passage through significant
material.

\section{Results}
\label{s:results}

\begin{figure}[!p]
        \vspace{0.8cm}
        \begin{center}
            \resizebox{\linewidth}{!}{\includegraphics{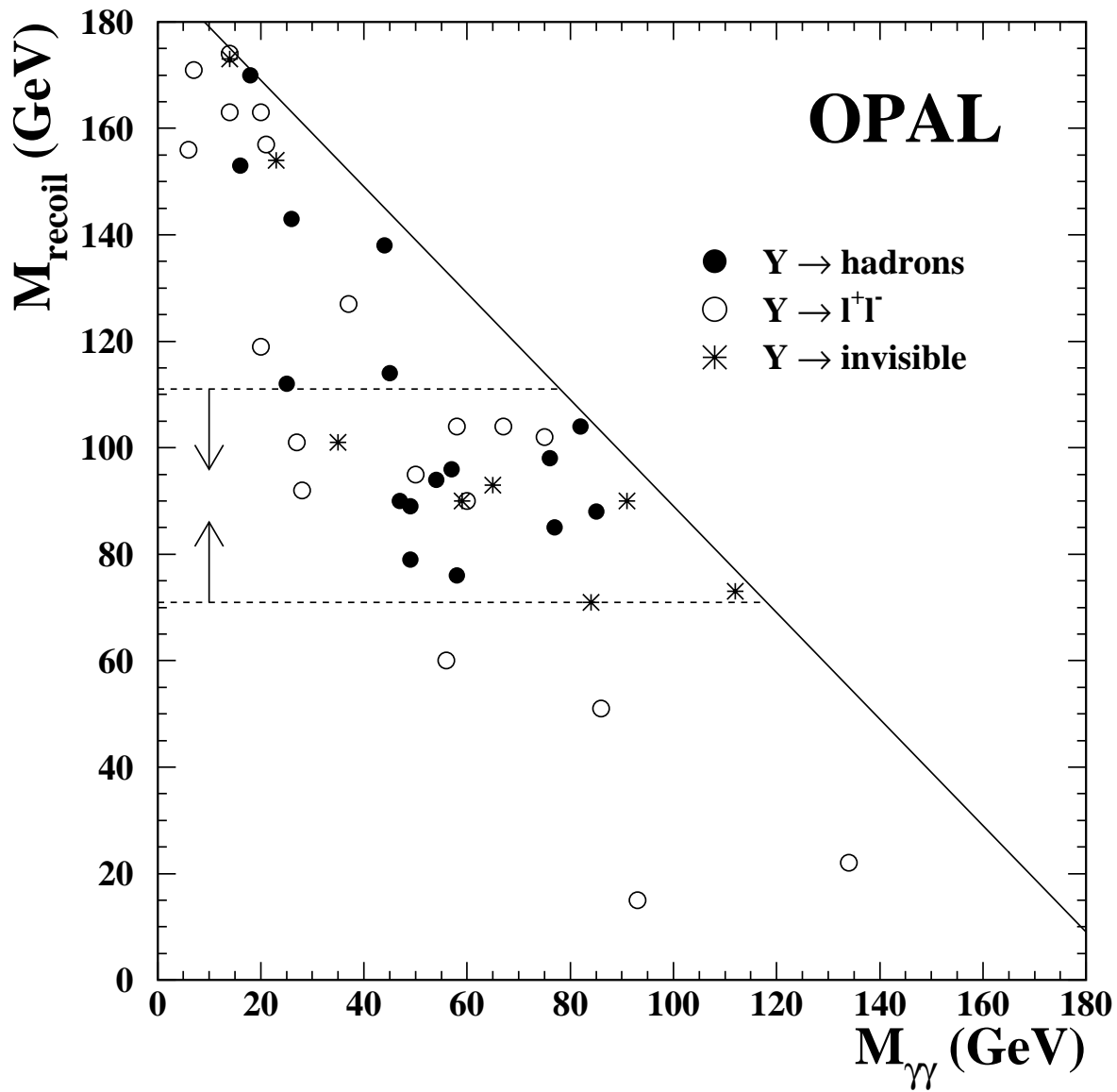} }
        \caption[COMGG2]{    
                  Distribution of mass recoiling against the di-photon system
                  versus di-photon invariant mass for events passing 
                  the general search cuts.
                  The different search channels are as indicated. 
                  The diagonal line denotes the kinematic limit.
                  Dashed lines and arrows indicate the events accepted
                  for the \Hzero\Zzero\ search.
        \label{COMGG2} }
        \end{center}
    \end{figure}

\begin{figure}[!p]
        \vspace{0.8cm}
        \begin{center}
           \resizebox{\linewidth}{!}{\includegraphics{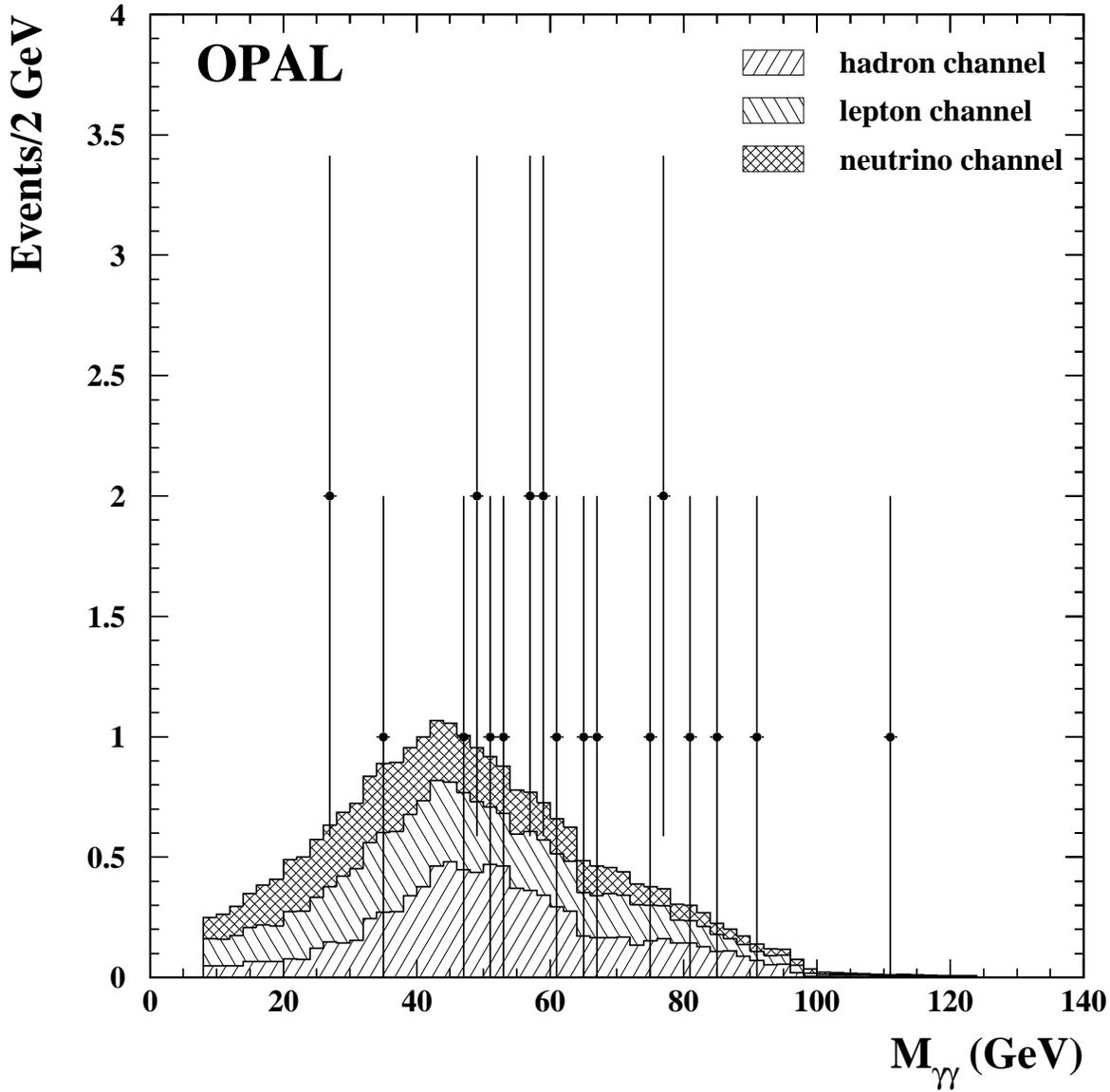} }
        \caption[COMGG]{    
                  Distribution of mass of the two highest-energy photons
                  in the \Hzero\Zzero\ search 
                  after application of all selection criteria.
                  All search channels are included.
                  Data are shown as points with error bars.
                  Background simulation is shown as a histogram 
                  showing the contributions from the hadronic, 
                  charged lepton and missing energy channels as
                  denoted.
        \label{COMGG} }
        \end{center}
    \end{figure}

Figure~\ref{COMGG2} shows the di-photon mass versus
the recoil mass for all candidate events passing the
general search cuts.  
The distribution of di-photon masses for the \Hzero\Zzero\ search candidates 
is shown in Figure~\ref{COMGG}, together with the 
simulation of Standard Model backgrounds.
Combining all three general search channels results in
44 observed events versus $54.2\pm2.5$ expected from Standard Model sources. 
Summing over all three \Hzero\Zzero\ channel decay modes and expected background sources 
yields $25.0\pm1.7$ events expected versus 22 observed.

\subsection{General Search Results}

Using only the data taken at \Ecm=189~GeV, we set
limits for the production mode $\epem \ra {\mrm XY}$, 
where X is any scalar resonance decaying into di-photons. 
The candidate events from the general search (no recoil mass cut) 
are used to set upper limits on 
$\sigma(\epem\ra {\mrm X Y})\times B({\mrm X} \ra \gaga)\times 
B(\mrm Y \ra f\bar{f})$.
Such results are valid
independent of the nature of Y, provided it decays to a fermion pair and has
negligible width.
The search is also restricted to X and Y masses above 10~GeV and below 180~GeV in order
to allow the decay products to have sufficient energies and momenta
to give reasonable search acceptances.

The event candidates from the general search are
used to calculate 95\% CL upper limits on the number
of events in 1~GeV [$\MX,\MY$] mass bins, 
where $\MX$ corresponds to the di-photon mass 
and $\MY$ to the recoil mass.
Efficiencies for signals were calculated using
two grids of simulated signals 
which were interpolated from
Monte Carlo samples generated in 10~GeV [$\MX,\MY$] steps using 
the $\epem \ra \Hboson \Zboson$ and
the $\epem \ra \Hboson \mrm A^0$ 
processes as models 
for the $\epem \ra {\mrm XY} \ra \gamma\gamma+\mrm f\bar{f}$ final state.
The grid was generated for X masses from 10 to 180~GeV and Y masses
from 10 to 180~GeV such that 
$\MX + \MY > M_{\mrm Z}$. 
This latter constraint was motivated by the higher sensitivity of searches performed
at \Ecm=$M_{\mrm Z}$.
For each [$\MX,\MY$] bin, the 95\% CL upper limit  
on the number of signal events is computed using the 
frequentist method of reference~\cite{JUNK}.
This statistical procedure incorporates the di-photon mass resolution
(typically less than 2~GeV for \mgg$<$100~GeV). 
The effect of the systematic error for efficiencies and background modelling
is incorporated by reducing the subtracted background by the systematic,
but using an additional systematic uncertainty of 5\%
to account for interpolation error in the efficiency grid 
(especially near kinematic limits).

\begin{figure}[!htb]
        \vspace{0.8cm}
        \begin{center}
            \resizebox{\linewidth}{!}{\includegraphics{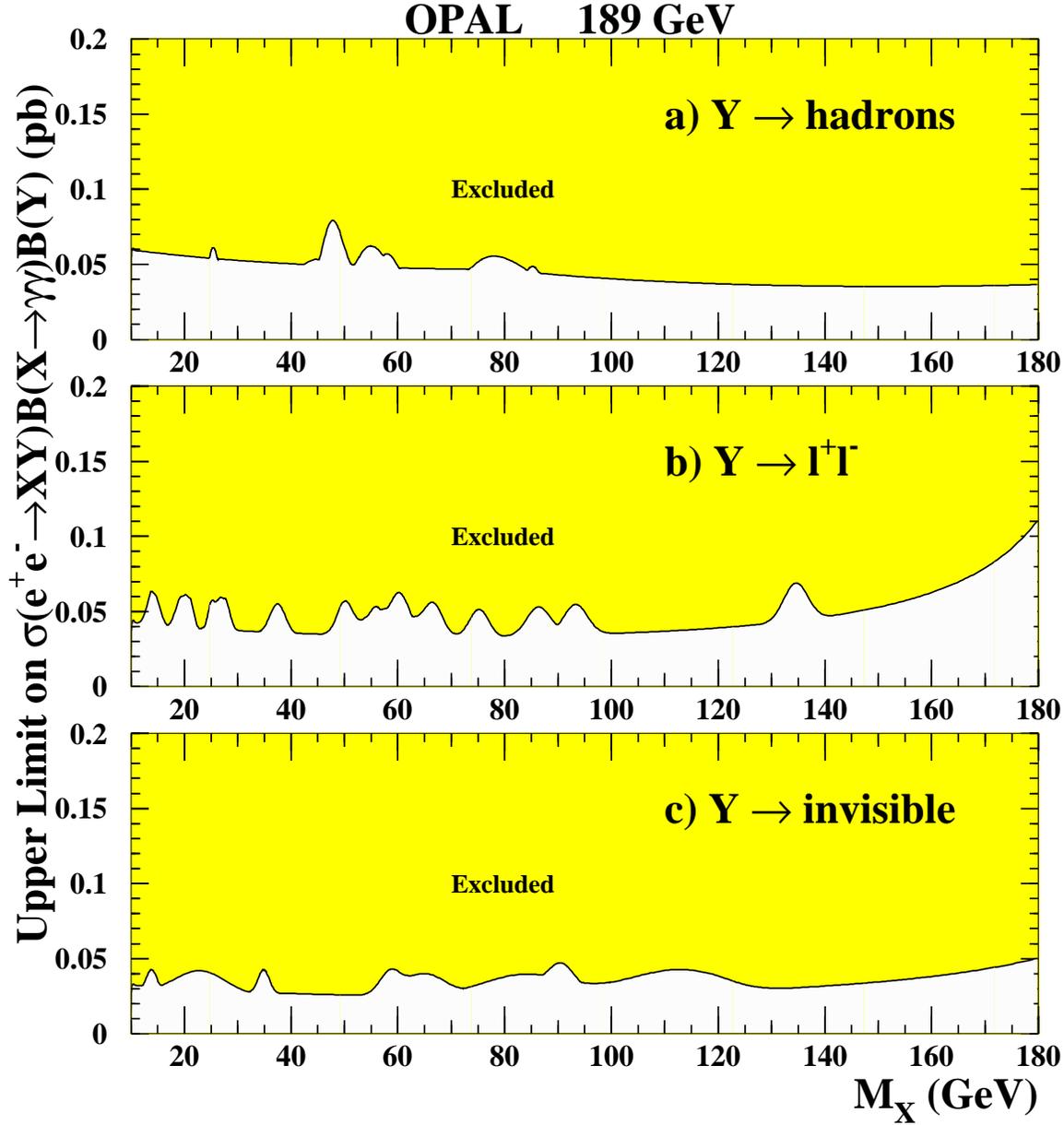} }
        \caption[limxy]{
                 95\% confidence level upper limit on
                 $\sigma(\epem\ra {\rm XY}) \times B(\mrm X \ra \gaga) \times B(Y)$
                 for the case 
                 where: a) Y decays hadronically, b) Y decays into any charged
                 lepton pair and c) Y decays invisibly. The limits
                 for each $\MX$ assume the smallest efficiency as a function of
                 $\MY$ such that $10 < \MY < 180$~GeV and that
                 $\MX + \MY > M_{\mrm Z}$.
        \label{limxy} }
        \end{center}
    \end{figure}

Figure~\ref{limxy} shows the 95\% CL upper limits on
$\sigma(\epem\ra {\mrm X Y})\times B({\mrm X} \ra \gaga)\times 
B(\mrm Y \ra f\bar{f})$.
To present the limits only as a function of $\MX$, the figure shows
the weakest limit obtained in each $\MX$ bin as $\MY$ was scanned
subject to the constraints mentioned above.
For a scalar/vector hypothesis for X/Y, the efficiency is found to be the
same to within 5\% with that for a scalar/scalar hypothesis;
the lower of these efficiencies is used in setting the limits.
For the lepton search channel, the efficiency for Y $\ra \tptm$ is used,
as it turns out to have the lowest of the dilepton efficiencies.
Cross section limits of 30 -- 100~fb are obtained 
over $10 < \MX < 180$~GeV.

\subsection{Search for the Standard Model Higgs Boson}

\begin{figure}[!htb]
        \vspace{0.8cm}
        \begin{center}
            \resizebox{\linewidth}{!}{\includegraphics{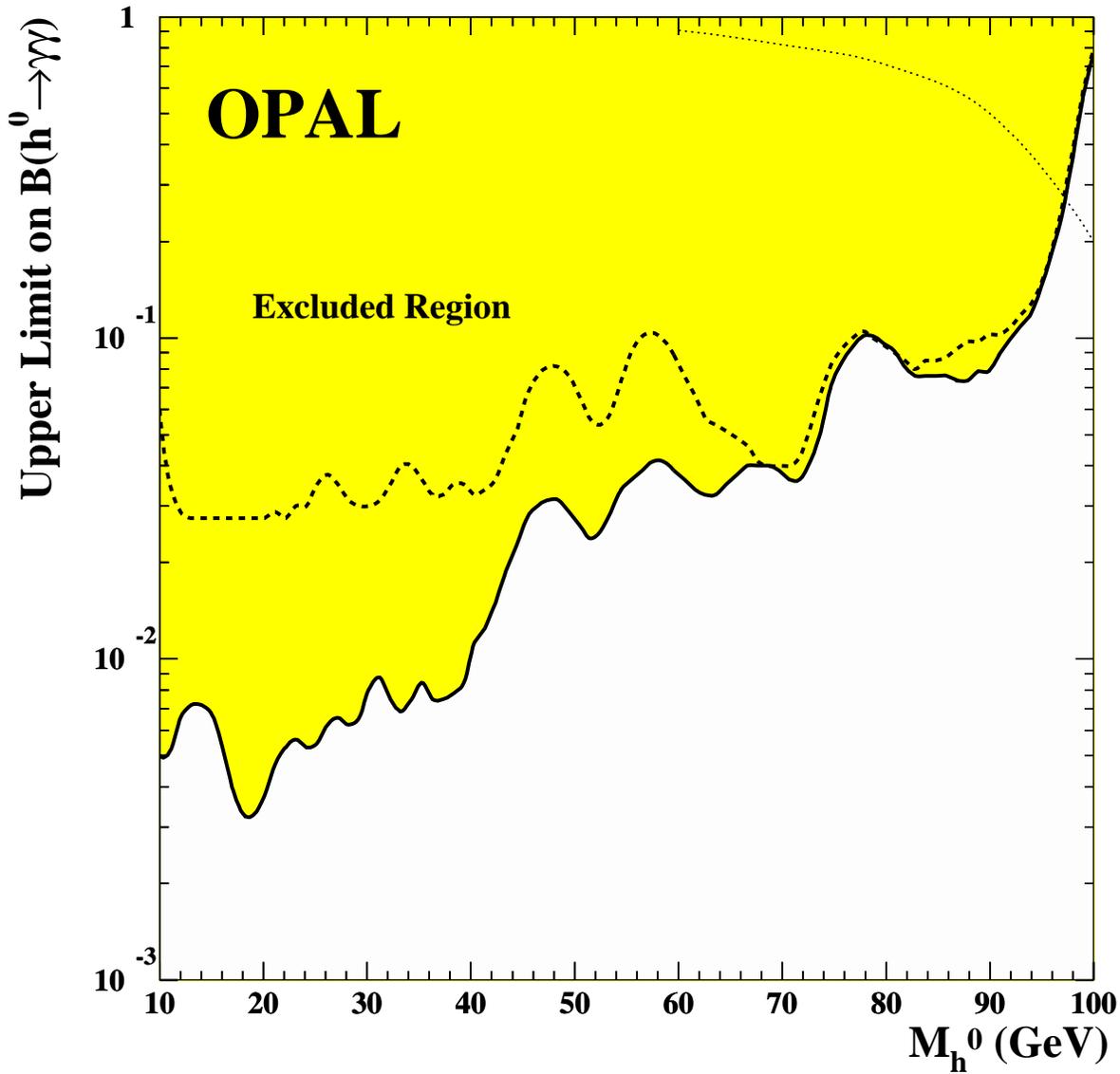} }
        \caption[bgglim]{
                 95\% confidence level upper limit on the branching fraction
                 $B$($\Hboson \ra \gaga$)
                 for a Standard Model Higgs boson production rate. 
                 The shaded region, obtained with all LEP energies, is excluded;
                 the dashed line shows the limit obtained with the 189~GeV data only. 
                 The dotted line
                 is the predicted $B$($\Hboson \ra \gaga$) assuming
                 $B$($\Hboson \ra \mrm f \bar{f}$)=0. 
                 The intersection of the dotted line with the exclusion curve gives a lower limit of
                 96.2~GeV for the fermiophobic Higgs model. 
        \label{bgglim} }
        \end{center}
    \end{figure}

The events passing all \Hzero\Zzero\ cuts are used to set an upper limit
on the di-photon branching ratio of a particle having the Standard Model
Higgs boson production rate.
For each 1~GeV di-photon mass bin, the 95\% CL upper limit  
on the number of signal events is computed using the 
frequentist method and background subtraction as in the previous section,
with the efficiencies now including the Standard Model \Zzero\ branching fractions.
Figure~\ref{bgglim} shows the 95\% CL upper limit for the
di-photon branching ratio 
obtained by combining the \Ecm=189~GeV candidate events
with those from OPAL searches at 
\Ecm=91--183~GeV~\cite{183paper,OPAL_HGG,OPAL_ggjj_1},
where the Standard Model $\Hboson \mrm Z^{0(*)}$ production cross section
is assumed at each \Ecm.
For masses lower than approximately 60~GeV, LEP-1 limits for $B(\Hboson \ra \gaga)$ 
have been inferred from references \cite{OPAL_ggjj_1,LowMgg}.

The limits on $B(\Hboson \ra \gaga)$ 
are used to rule out Higgs bosons in certain 
non-minimal models. 
Shown in Figure~\ref{bgglim} is the $\Hboson \ra \gaga$ branching
ratio in the Standard Model
computed using HDECAY~\cite{HDECAY} 
with the fermionic couplings switched off.
A 95\% CL lower mass limit for such fermiophobic Higgs bosons is set
at 96.2~GeV, where the predicted branching ratio crosses the upper-limit curve.

\subsection{The Higgs Triplet Model}

It is possible that a non-minimal Higgs sector
incorporates triplet fields; particles formed exclusively
from such fields are fermiophobic. 
The minimal Higgs Triplet model (HTM)~\cite{Triplets,Akeroyd}
requires the inclusion of two triplet fields in order to have
the $\rho$-parameter near unity. The model has 
10 Higgs bosons in the form of
a fiveplet ($H_{5}$), a threeplet ($H_{3}$), and two singlets ($H_{1}$).
The $H_{5}^{0}$ and one of the singlets, $H_{1}^{0'}$, are formed from the
triplet field, apart from possible mixing with doublet components.
Akeroyd~\cite{Akeroyd} has shown that measurements constrain the
mixing parameters so that the $H_{1}^{0'}$ is almost entirely fermiophobic,
and therefore could be interpreted as the X in this search.
 
The process 
$\epem \ra \mrm H_{1}^{0'}\Zboson$ occurs at the Standard Model \Hzero\Zzero\
rate modified by the factor $\frac{8}{3}\sin^2\!\theta_H$, where
the angle $\theta_H$ is a parameter of the model describing
the mixing of the doublet and triplet fields. 
Limits on $\theta_H$ can therefore be inferred from Figure~\ref{bgglim}
by dividing the upper limit by the fermiophobic di-photon branching ratio.
The limits on $\theta_H$ obtained from this experiment are more restrictive
than limits inferred from the \Zzero\ width~\cite{Akeroyd} up to 
an $H_{1}^{0'}$ mass of approximately 96~GeV.

\section{Conclusions}

A  search for the production of Higgs bosons and other
new particles decaying to photon pairs has been performed 
using 182.6~\ipb\ of data taken at an average 
centre-of-mass energy of 188.6~GeV. 
Model independent upper limits are obtained on
$\sigma(\epem\ra {\mrm X Y})\times B({\mrm X} \ra \gaga)\times B(\mrm Y \ra f\bar{f})$.
Limits of 30 -- 100~fb are obtained over $10 < \MX < 180$~GeV,
where $10 < \MY < 180$~GeV and $\MX + \MY > M_{\mrm Z}$,
for Y either a scalar or vector particle, 
provided that the Y decays to a fermion pair.

The results of this search have been
combined with previous OPAL results to set limits on $B$($\Hboson \ra \gaga$) 
up to a Higgs boson mass of 100~GeV, 
provided the Higgs particle is produced 
via $\epem \ra \Hboson \Zboson$ at the Standard Model rate.
A lower mass bound of 96.2~GeV is set at the 95\% confidence level for
Higgs particles which do not couple to fermions. 
%
%
\bigskip\bigskip\bigskip
\begin{flushleft}
{\Large\bf Acknowledgements}
\end{flushleft}
\par
We particularly wish to thank the SL Division for the efficient operation
of the LEP accelerator at all energies
and for their continuing close cooperation with
our experimental group.  We thank our colleagues from CEA, DAPNIA/SPP,
CE-Saclay for their efforts over the years on the time-of-flight and trigger
systems which we continue to use.  In addition to the support staff at our own
institutions we are pleased to acknowledge the  \\
Department of Energy, USA, \\
National Science Foundation, USA, \\
Particle Physics and Astronomy Research Council, UK, \\
Natural Sciences and Engineering Research Council, Canada, \\
Israel Science Foundation, administered by the Israel
Academy of Science and Humanities, \\
Minerva Gesellschaft, \\
Benoziyo Center for High Energy Physics,\\
Japanese Ministry of Education, Science and Culture (the
Monbusho) and a grant under the Monbusho International
Science Research Program,\\
Japanese Society for the Promotion of Science (JSPS),\\
German Israeli Bi-national Science Foundation (GIF), \\
Bundesministerium f\"ur Bildung, Wissenschaft,
Forschung und Technologie, Germany, \\
National Research Council of Canada, \\
Research Corporation, USA,\\
Hungarian Foundation for Scientific Research, OTKA T-016660, 
T023793 and OTKA F-023259.\\

\bigskip\bigskip\bigskip\bigskip\bigskip\bigskip

\newpage


\def\mulc1{\multicolumn{1}{|c|}}

\begin{table}[htbp]
\begin{center}
\begin{tabular}{|l||r||r|r|r|}\hline
 Cut  & Data & \mulc1{$\Sigma$Bkgd} &\mulc1{$(\gamma/{\rm Z})^{\ast}$} & \mulc1{4f} \\ 
\hline\hline
 (A1)    & 10075 & 10024.6  &  7321.0  &  2703.6  \\ \hline
 (A2)    & 63    & 54.0     &  51.6    &  2.4     \\ \hline
 (A3)    & 38    & 41.4     &  39.7    &  1.7     \\ \hline
 (A4)    & 16    & $17.4\pm1.7$    & 15.9  & 1.5  \\ \hline \hline
 (A5)    & 10    & $9.0\pm1.3$     & 8.7  & 0.3 \\ \hline 
\end{tabular}
\end{center}
  \caption[Data and MC after cuts]
  {Events remaining in the hadronic search channel after the indicated cumulative cuts
   described in Section \ref{s:qqgg}. 
   The entry for (A4) is used in the general search. 
   The entry for (A5) is for the $\Mrec$ cut for the \Hzero\Zzero\ search. 
   In addition to the total simulated background ($\Sigma$Bkgd), the components from
   $(\gamma/{\rm Z})^{\ast}$ and four-fermion (``4f") final states
   are shown. 
}
  \label{T:qq1}

\end{table}


\begin{table}[!htbp]                      
\begin{center}
\begin{tabular}{|l||r||r|r|r|r|r|r|}\hline
 Cut     & Data   &\mulc1{$\Sigma$Bkgd} &\mulc1{$\epem$} & \mulc1{$\tptm$}  
                                        & \mulc1{$\mpmm$} & \mulc1{$\gaga$}  
                                                             & \mulc1{\eeff}  \\ 
\hline\hline
 (B1)  & 41115 & 36126.6 & 34679.8 & 646.9 & 36.3 & 281.1 & 482.4   \\ \hline
 (B2)  & 159   & 168.4   & 66.8    & 8.8   & 5.5  & 86.1  & 1.1     \\ \hline
 (B3)  & 146   & 161.6   & 62.1    & 8.3   & 5.2  & 84.9  & 1.1     \\ \hline
 (B4)  & 20    &$25.6\pm1.8$ & 16.2    & 3.9   & 5.0  & 0.2   & 0.3     \\ \hline \hline
 (B5)  & 7     &$8.9\pm1.0$  & 4.7  & 1.5    & 2.5   & 0.0    & 0.1   \\ \hline
\end{tabular}
\end{center}
  \caption[Data and MC after cuts]
  {Events remaining for the leptonic channel analysis after the indicated 
   cumulative cuts described in Section \ref{s:llgg}.
   In addition to the total simulated background ($\Sigma$Bkgd), 
   the individual contributions from Bhabha scattering ($\epem$),
   $\tau$-pair, $\mu$-pair, $\gaga$ and \eeff\ final states 
   are shown. 
   Criterion (B5), the recoil mass cut, is only applied for the \Hzero\Zzero\ search. 
}  \label{T:ggll}
\end{table}

%

\begin{table}[htbp]
\begin{center}
\begin{tabular}{|l||r||r|r|r|r|r|r|}\hline
 Cut  & Data & \mulc1{$\Sigma$Bkgd} & \mulc1{$\nunu\gaga$} & \mulc1{$\gaga$} 
                     & \mulc1{$\epem$} & \mulc1{$\ell^+\ell^-$} & \mulc1{\eeff}  \\ \hline\hline
 (C1)   & 213061 & 118393.4 & 40.9 & 2809.8 & 114303.6 & 141.3 & 1097.8   \\ \hline
 (C2)   & 323    & 287.9    & 11.3 & 232.7  & 42.9     & 0.7   & 0.3      \\ \hline
 (C3)   & 70     & 64.3     & 11.0 & 26.5   & 26.0     & 0.6   & 0.2      \\ \hline
 (C4)   & 34     & 36.1     & 10.8 & 24.6   & 0.6      & 0.0   & 0.1      \\ \hline
 (C5)   & 8      &$11.2\pm0.5$& 10.4 & 0.3    & 0.4      & 0.0   & 0.1      \\ \hline \hline
 (C6)   & 5      &$7.1\pm0.3$ & 7.0 & 0.0 & 0.0  & 0.0 & 0.1 \\ \hline
\end{tabular}
\end{center}

  \caption[Data and MC after cuts]
  {Events remaining after the indicated cumulative cuts for the 
   missing energy search channel described in Section \ref{s:nngg}. 
   In addition to the total simulated background ($\Sigma$Bkgd), 
   the individual contributions from 
   $\nunu\gaga$, $\gaga$, $\epem$-pair, lepton pair ($\ell\equiv\mu,\tau$) 
   production and \eeff\ final states are shown. 
   Criterion (C6), the recoil mass cut, is only applied for the \Hzero\Zzero\ search.
}  \label{T:ggnn}
\end{table}


\begin{table}[htbp]
\begin{center}
 \begin{tabular}{|l||c|c|c|c|c||c|c|c|c|c|}
\cline{2-11}
\multicolumn{1}{c}{ }  & \multicolumn{10}{|c|}{Efficiency (\%)} \\ 
\cline{2-11}
\multicolumn{1}{c}{ }  & \multicolumn{5}{|c|}{General Search}
& \multicolumn{5}{|c|}{\Hzero\Zzero\ Search} \\ 
\hline
\mgg\ (GeV):    & 30 & 50 & 70 & 90 & 100   & 30 & 50 & 70 & 90 & 100   \\ \hline\hline
$\qqbar\gaga$   & 35 & 37 & 39 & 39 & 38    & 29 & 35 & 46 & 57 & 49  \\ \hline
$\ell\ell\gaga$ & 45 & 49 & 50 & 48 & 47    & 44 & 49 & 54 & 56 & 44  \\ \hline
$\nn\gaga$      & 59 & 65 & 67 & 66 & 64    & 49 & 57 & 59 & 67 & 48  \\ \hline 

\end{tabular}
\end{center}
\caption[Acceptances for \Hzero\Zzero\ search channel.]
{Efficiency in percent (\%) for each \Hzero\Zzero\ and general search channel for Higgs masses
as indicated. 
The general search numbers indicate the minimum efficiency for
variation of the recoil mass $\MY$ over 
$\mrm{E_{cm}} > \MX + \MY > M_{\mrm Z}$,
where $\MX$ is the di-photon mass; 
the minimum general efficiency can therefore be smaller than the \Hzero\Zzero\ efficiency.
}
\label{t:hzeff}
\end{table}

\newpage
\newpage

\newpage

\newpage

\newpage

\end{document}